\documentclass[namedreferences]{spackap}

\usepackage{graphicx} 
\usepackage{times} 
\def\Yi{\eta^{\ast}_{11} \left( \frac{y_{i}}{2} g' Z_{\chi 1} + 
        g T_{3i} Z_{\chi 2} \right) + \eta^{\ast}_{12} 
        \frac{g m_{q_{i}} Z_{\chi 5-i}}{2 m_{W} B_{i}}}

\def\Xii{\eta^{\ast}_{11} 
        \frac{g m_{q_{i}}Z_{\chi 5-i}^{\ast}}{2 m_{W} B_{i}} - 
        \eta_{12}^{\ast} e_{i} g' Z_{\chi 1}^{\ast}}

\def\Wi{\eta_{21}^{\ast}
        \frac{g m_{q_{i}}Z_{\chi 5-i}^{\ast}}{2 m_{W} B_{i}} -
        \eta_{22}^{\ast} e_{i} g' Z_{\chi 1}^{\ast}}

\def\Vi{\eta_{22}^{\ast} \frac{g m_{q_{i}} Z_{\chi 5-i}}{2 m_{W} B_{i}}
        + \eta_{21}^{\ast}\left( \frac{y_{i}}{2} g' Z_{\chi 1}
        + g T_{3i} Z_{\chi 2} \right)}

\def\zthree{\delta_{1i} [g Z_{\chi 2} - g' Z_{\chi 1}]}

\def\zfour{\delta_{2i} [g Z_{\chi 2} - g' Z_{\chi 1}]}
\def\beq{\begin{equation}}
\def\eeq{\end{equation}}
\def\ga{\mathrel{\raise.3ex\hbox{$>$\kern-.75em\lower1ex\hbox{$\sim$}}}}
\def\la{\mathrel{\raise.3ex\hbox{$<$\kern-.75em\lower1ex\hbox{$\sim$}}}}

\begin{document}

\begin{article}

\begin{opening}
\title{Theoretical Aspects of Dark Matter Detection}
\runningtitle{Theoretical Aspects of Dark Matter Detection}

\author{John \surname{Ellis}\email{john.ellis@cern.ch}}
\institute{TH Division, CERN, Geneva, Switzerland}
\author{Andrew \surname{Ferstl}\email{Andrew.Ferstl@winona.msus.edu}}
\institute{Department of Physics, Winona State University, Winona, MN USA}
\author{Keith A. \surname{Olive}\email{olive@umn.edu}}
\institute{TH Division, CERN, Geneva, Switzerland \\
and \\
Theoretical Physics Institute,
University of Minnesota, Minneapolis, MN 55455, USA}
\runningauthor{J. Ellis, A. Ferstl, and K.A. Olive}

\date{Received \ldots, accepted \ldots}

\begin{ao}
Keith A. Olive,
Theoretical Physics Institute,
University of Minnesota, Minneapolis, MN 55455, USA
olive@umn.edu
\end{ao}

\begin{abstract}
Direct and indirect dark matter detection relies on the scattering
of the dark matter candidate on nucleons or nuclei. Here, attention is 
focused on dark matter candidates (neutralinos) predicted in the minimal supersymmetric
standard model and its constrained version with universal input soft
supersymmetry-breaking masses.
Current expectations for elastic scattering cross sections for neutralinos on protons are
discussed with particular attention to satisfying all current accelerator constraints as
well as insuring a sufficient cosmological relic density to account for 
the dark matter in the Universe. 
\end{abstract}

\end{opening}

\vskip -5in
\rightline{hep-ph/0106148}
\rightline{CERN--TH/2001-153}
\rightline{UMN--TH--2012/01, TPI--MINN--01/28}
\vskip 4in

\section{Introduction}
\label{sec:introduction}
Minimal supersymmetric theories with R-parity conservation are particularly
attractive for the study of dark matter as they predict the existence of 
a new stable particle which is the lightest R-odd state (the LSP).  Furthermore, for
parameter values of interest to resolve the gauge hierarchy problem, the LSP
has an annihilation cross section which yields a relic density of cosmological interest
\cite{EHNOS}.  Recent accelerator constraints have made a great impact on
the available parameter space in the MSSM and in particular the constrained version
or CMSSM in which all scalar masses are assumed to be unified at a grand unified scale
\cite{EFGOS,EFGO,EGNO,ENO}.  In addition, significant progress has been made concerning
the relic density calculations \cite{EFOSi,EFGOSi,GLP}. These issues were 
discussed in detail in John Ellis' contribution and will only be touched on briefly here.

The main impact of the null searches, particularly at LEP, is in the increase
in the lower limit to the LSP mass as well as the rest of the sparticle
spectrum.
This has the unfortunate effect of lowering the elastic scattering cross sections for
neutralinos on protons, making direct detection experiments more difficult.
Nevertheless, we have now entered a period where accelerator constraints will be at a
lull due to the transition from LEP to the LHC and the time required for run II at the
Tevatron to acquire the needed luminosity\footnote{We can however, expect
improvements in
the uncertainties in rare B decays and the measurement of the anomalous magnetic moment of
the muon, both of which will have an impact on the allowed
supersymmetric parameter space.}.
Therefore, during the next several years we can be hopeful that direct detection
experiments will improve to make meaningful inroads to the
supersymmetric dark matter parameter
space.  

In this contribution, we will discuss the current status of the expected 
neutralino-proton elastic cross sections in the MSSM and CMSSM.  These results will be
applied to the possibility of direct dark matter detection. 

As noted earlier, we will restrict our attention to regions of parameter space for which
the relic density of neutralinos is of cosmological interest.
For an age of the Universe $t > 12$ Gyr, there is a firm upper bound on the
relic density $\Omega_\chi h^2 < 0.3$, where $\Omega_\chi$ is the fraction of 
critical density in the form of neutralinos, $\chi$, and $h$ is the Hubble parameter
in units of 100 km/s/Mpc.  This limit represents a strict cosmological bound on the
supersymmetric parameter space. We also focus our discussion on the
parameter values which
lead to relic densities with $\Omega_\chi h^2 > 0.1$.  While this does not place
any bound on supersymmetry, it is a reasonable requirement for dark matter candidates.
Neutralinos with a lower density could not be the dominant form of dark matter in our
Galaxy, and therefore detection rates would necessarily be suppressed. 

\section{The MSSM vs. The CMSSM}
\label{sec:streamers}

As discussed here by John Ellis, 
the neutralino LSP is the
lowest-mass eigenstate combination of the Bino ${\tilde B}$, Wino $\tilde W$
and Higgsinos ${\tilde H}_{1,2}$, whose mass matrix $N$ is
diagonalized by a matrix $Z$: $diag(m_{\chi_{1,..,4}}) = Z^* N Z^{-1}$.
The composition of the lightest neutralino may be written as
\begin{equation}
\chi = Z_{\chi 1}\tilde{B} + Z_{\chi 2}\tilde{W} +
Z_{\chi 3}\tilde{H_{1}} + Z_{\chi 4}\tilde{H_{2}}
\label{id}
\end{equation}
 We assume
universality at the supersymmetric GUT scale for the  gauge couplings as well as
gaugino masses: $M_{1,2,3} = m_{1/2}$, so that 
$M_1 \sim \frac{5}{3}\tan^2{\theta_{W}}M_{2}$ at the electroweak scale
(note that this relation is not exact when two-loop running of gauge
sector is included, as done here).

We also assume
GUT-scale universality for the soft supersymmetry-breaking scalar masses
$m_0$ of the squarks and sleptons.  In the case of the CMSSM, the universality is
extended to 
the soft masses of the Higgs bosons as well.
We further
assume GUT-scale universality for the soft supersymmetry-breaking
trilinear terms $A_0$.
In the MSSM,
we treat as free parameters $m_{1/2}$ (we actually use $M_2$ which is
equal to $m_{1/2}$ at the unification scale), the soft
supersymmetry-breaking scalar mass scale
$m_{0}$ (which in the present context refers only to the universal
sfermion masses at the unification scale),
$A_0$ and $\tan\beta$. In addition, we treat $\mu$ and the pseudoscalar
Higgs mass
$m_A$ as independent parameters, and thus the two Higgs soft
masses $m_1$ and $m_2$, are specified by the electroweak vacuum
conditions, which we calculate using $m_t = 175$~GeV. 
In contrast, in the CMSSM,   $m_1$ and $m_2$ are set equal to 
$m_0$ at the GUT scale and hence $\mu$ (up to a sign) and $m_A$ are calculated
quantities, their values being fixed by the electroweak symmetry breaking conditions.

%
%

\section{Elastic Scattering Cross Sections}
\label{sec:coronalholes}

The MSSM Lagrangian leads to the following low-energy effective
four-fermion
Lagrangian suitable for describing elastic $\chi$-nucleon
scattering~\cite{FFO1}:
\begin{eqnarray}
{\cal L} & = & \bar{\chi} \gamma^\mu \gamma^5 \chi \bar{q_{i}} 
\gamma_{\mu} (\alpha_{1i} + \alpha_{2i} \gamma^{5}) q_{i} +
\alpha_{3i} \bar{\chi} \chi \bar{q_{i}} q_{i} \nonumber \\
& + & \alpha_{4i} \bar{\chi} \gamma^{5} \chi \bar{q_{i}} \gamma^{5} q_{i}+
\alpha_{5i} \bar{\chi} \chi \bar{q_{i}} \gamma^{5} q_{i} +
\alpha_{6i} \bar{\chi} \gamma^{5} \chi \bar{q_{i}} q_{i}
\label{lagr}
\end{eqnarray}
This Lagrangian is to be summed over the quark generations, and the 
subscript $i$ labels up-type quarks ($i=1$) and down-type quarks
($i=2$).  The terms with coefficients $\alpha_{1i}, \alpha_{4i},
\alpha_{5i}$ and $\alpha_{6i}$ make contributions to the 
elastic scattering cross section that are velocity-dependent,
and may be neglected for our purposes. In fact, if the 
CP-violating phases are absent as assumed here, $\alpha_5 = \alpha_6 =
0$~\cite{FFO2,CIN}. The coefficients relevant for our discussion are the 
spin-dependent coefficient, $\alpha_2$, 

\begin{eqnarray}
\alpha_{2i} & = & \frac{1}{4(m^{2}_{1i} - m^{2}_{\chi})} \left[
\left| Y_{i} \right|^{2} + \left| X_{i} \right|^{2} \right] 
+ \frac{1}{4(m^{2}_{2i} - m^{2}_{\chi})} \left[ 
\left| V_{i} \right|^{2} + \left| W_{i} \right|^{2} \right] \nonumber \\
& & \mbox{} - \frac{g^{2}}{4 m_{Z}^{2} \cos^{2}{\theta_{W}}} \left[
\left| Z_{\chi_{3}} \right|^{2} - \left| Z_{\chi_{4}} \right|^{2}
\right] \frac{T_{3i}}{2}
\label{alpha2}
\end{eqnarray}
and the spin-independent or scalar coefficient, $\alpha_3$,
\begin{eqnarray}
\alpha_{3i} & = & - \frac{1}{2(m^{2}_{1i} - m^{2}_{\chi})} Re \left[
\left( X_{i} \right) \left( Y_{i} \right)^{\ast} \right] 
- \frac{1}{2(m^{2}_{2i} - m^{2}_{\chi})} Re \left[ 
\left( W_{i} \right) \left( V_{i} \right)^{\ast} \right] \nonumber \\
& & \mbox{} - \frac{g m_{qi}}{4 m_{W} B_{i}} \left[ Re \left( 
\zthree \right) D_{i} C_{i} \left( - \frac{1}{m^{2}_{H_{1}}} + 
\frac{1}{m^{2}_{H_{2}}} \right) \right. \nonumber \\
& & \mbox{} +  Re \left. \left( \zfour \right) \left( 
\frac{D_{i}^{2}}{m^{2}_{H_{2}}}+ \frac{C_{i}^{2}}{m^{2}_{H_{1}}} 
\right) \right]
\label{alpha3}
\end{eqnarray}
where
\begin{eqnarray}
X_{i}& \equiv& \Xii \nonumber \\
Y_{i}& \equiv& \Yi \nonumber \\
W_{i}& \equiv& \Wi \nonumber \\
V_{i}& \equiv& \Vi
\label{xywz}
\end{eqnarray}
where $y_i, T_{3i}$ denote hypercharge and isospin, and
\begin{eqnarray}
\delta_{1i} = Z_{\chi 3} (Z_{\chi 4}) &,& \delta_{2i} = Z_{\chi 4}
(-Z_{\chi 3}), \nonumber \\
B_{i} = \sin{\beta} (\cos{\beta}) &,& A_{i} = \cos{\beta} ( -\sin{\beta}), 
\nonumber \\
C_{i} = \sin{\alpha} (\cos{\alpha}) &,& D_{i} = \cos{\alpha} ( -
\sin{\alpha}) 
\label{moredefs}
\end{eqnarray}
for up (down) type quarks. We denote by $m_{H_2} < m_{H_1}$
the two scalar Higgs masses, and $ \alpha $ denotes the Higgs mixing
angle. Finally, the sfermion mass-squared matrix is diagonalized by a
matrix
$\eta$: $diag(m^2_1, m^2_2) \equiv
\eta M^2 \eta^{-1}$, which can be parameterized 
for each flavour $f$ by an angle $ \theta_{f} $:
\begin{equation}
\left( \begin{array}{cc}
\cos{\theta_{f}} & \sin{\theta_{f}}  \nonumber \\
-\sin{\theta_{f}} & \cos{\theta_{f}}
\end{array} \right) 
\hspace{0.5cm}
 \equiv 
\hspace{0.5cm}
 \left( \begin{array}{cc}
\eta_{11} & \eta_{12} \nonumber \\
\eta_{21} & \eta_{22}
\end{array} \right)
\label{defineeta}
\end{equation}

The spin-dependent part of the elastic $\chi$-nucleus cross section can be
written as
\begin{equation}
\sigma_{2} = \frac{32}{\pi} G_{F}^{2} m_{r}^{2} \Lambda^{2} J(J + 1)
\label{sd}
\end{equation}
where $m_{r}$ is again the reduced neutralino mass, $J$ is the spin 
of the nucleus, and
\begin{equation}
\Lambda \equiv \frac{1}{J} (a_{p} \langle S_{p} \rangle + a_{n} \langle
S_{n} \rangle)
\label{lamda}
\end{equation}
where
\begin{equation}
a_{p} = \sum_{i} \frac{\alpha_{2i}}{\sqrt{2} G_{f}} \Delta_{i}^{(p)}, 
a_{n} = \sum_{i} \frac{\alpha_{2i}}{\sqrt{2} G_{f}} \Delta_{i}^{(n)}
\label{a}
\end{equation}
The factors $\Delta_{i}^{(p,n)}$ parameterize the quark spin content of the
nucleon. A recent global analysis of QCD sum rules for the $g_1$
structure functions~\cite{Mallot}, including ${\cal O}(\alpha_s^3)$
corrections,
corresponds formally to the values

\begin{eqnarray}
\Delta_{u}^{(p)} = 0.78 \pm 0.02,~~ & \Delta_{d}^{(p)} = -0.48 \pm
0.02 \nonumber \\
 \Delta_{s}^{(p)} = - 0.15 \pm 0.02 &
\label{spincontent}
\end{eqnarray}

The scalar part of the
cross section can be written as
\begin{equation}
\sigma_{3} = \frac{4 m_{r}^{2}}{\pi} \left[ Z f_{p} + (A-Z) f_{n} 
\right]^{2}
\label{si}
\end{equation}
where 
\begin{equation}
\frac{f_{p}}{m_{p}} = \sum_{q=u,d,s} f_{Tq}^{(p)} 
\frac{\alpha_{3q}}{m_{q}} +
\frac{2}{27} f_{TG}^{(p)} \sum_{c,b,t} \frac{\alpha_{3q}}{m_q}
\label{f}
\end{equation}
and $f_{n}$ has a similar expression.  The parameters
$f_{Tq}^{(p)}$  are defined
by
\begin{equation}
m_p f_{Tq}^{(p)} \equiv \langle p | m_{q} \bar{q} q | p \rangle
\equiv m_q B_q
\label{defbq}
\end{equation}
whilst $ f_{TG}^{(p)} = 1 - \sum_{q=u,d,s} f_{Tq}^{(p)} $~\cite{SVZ}.
We observe that only the products $m_q B_q$, the ratios of the quark masses
$m_q$ and the ratios of the scalar matrix elements $B_q$ are
invariant under renormalization and hence physical quantities.

We take the ratios of the quark masses from~\cite{leut}:
\beq
{m_u \over m_d} = 0.553 \pm 0.043 , \qquad
{m_s \over m_d} = 18.9 \pm 0.8
\eeq
In order to determine the ratios of the $B_q$ and the products $m_q B_q$
we use information from chiral symmetry applied to baryons.
Following~\cite{Cheng}, we have:
\beq
z \equiv {B_u - B_s \over B_d - B_s} =
{m_{\Xi^0} + m_{\Xi^-} -m_p -m_n \over 
m_{\Sigma^+} + m_{\Sigma^-} -m_p -m_n}
\label{cheng}
\eeq
Substituting the experimental values of these baryon masses, we find
\beq
z = 1.49
\label{chengvalue}
\eeq
with an experimental error that is negligible compared with others
discussed below. Defining
\beq
y \equiv {2 B_s \over B_d + B_u},
\label{defy}
\eeq
we then have 
\beq
{B_d \over B_u} = {2 + ((z - 1) \times y) \over 2 \times z - ((z - 1) \times
y)}
\label{bdoverbu}
\eeq
The experimental value of the $\pi$-nucleon $\sigma$ term
is~\cite{Gasser,knecht}:
\beq
\sigma \equiv {1 \over 2} (m_u + m_d) \times (B_d + B_u) = 45 \pm 8~{\rm MeV}
\label{sigma}
\eeq 
and octet baryon mass differences may be used to estimate
that~\cite{Gasser,knecht}
\beq
\sigma = {\sigma_0 \over (1 - y)}: \qquad \sigma_0 = 36 \pm 7~{\rm MeV}
\label{sigma0}
\eeq
The larger value of $\sigma = 65$ MeV \cite{ol,pav} considered by \cite{arn2}
leads to scattering cross section which are larger by a factor of  about 3.
Comparing (\ref{sigma}) and (\ref{sigma0}), we find a central value of
$y = 0.2$, to which we assign an error $\pm 0.1$, yielding
\beq
{B_d \over B_u} = 0.73 \pm 0.02
\eeq
The formal error in $y$ derived from (\ref{sigma}) and (\ref{sigma0}) is
actually $\pm$0.2, which would double the error in $B_d/B_u$. We have
chosen the smaller uncertainty because we consider a value
of y in excess of 30\% rather unlikely. 

The numerical magnitudes of the
individual renormalization-invariant products
$m_q B_q$ and hence the
$f_{Tq}^{(p)}$ may now be determined:
\begin{eqnarray}
f_{Tu}^{(p)} = 0.020 \pm 0.004, & f_{Td}^{(p)} = 0.026 \pm 0.005
\nonumber \\
\qquad f_{Ts}^{(p)} = 0.118 \pm 0.062 &
\label{pinput}
\end{eqnarray}
where essentially all the error in $f_{Ts}^{(p)}$ arises from the
uncertainty in $y$. The corresponding values for the neutron are
\begin{eqnarray}
f_{Tu}^{(n)} = 0.014 \pm 0.003, & f_{Td}^{(n)} = 0.036 \pm 0.008
\nonumber \\
\qquad f_{Ts}^{(n)} = 0.118 \pm 0.062 &
\end{eqnarray}
It is clear already that the difference between the scalar parts
of the cross sections for scattering off protons and neutrons must be
rather small.

\section{Results}
\label{sec:boundary}
We begin by discussing the results for the CMSSM \cite{EFlO1}. For fixed $\tan \beta$ and 
sign of $\mu$, we scan over experimentally and cosmologically allowed regions in the 
$m_{1/2} - m_0$ plane. Results here are shown for $A_0=0$. The combination of the
cosmological constraint $\Omega h^2 < 0.3$ and the constraint from the Higgs mass,
$m_{H_2} > 113$ GeV, eliminates low values of $\tan \beta \la 5$
\cite{EGNO}.  For the value of $\tan \beta = 10$, we show in Figure 1 the elastic
scattering cross section for spin-dependent (a,b) and scalar (c,d)
processes as
a function of the neutralino mass. 
Although it
is barely discernible, the thicknesses of the central curves in the
panels show the ranges in the cross section for fixed $m_\chi$
that are induced by varying $m_0$. At large
$m_\chi$ where coannihilations are important, the range in the allowed
values of $m_0$ is small and 
particularly little variation in the cross section is
expected.  The shaded regions show the effects of the
uncertainties in the input values of the $\Delta^{(p)}_i$ 
(\ref{spincontent}) (a,b) and in the $f_{T}^{(p)}$ (\ref{pinput}) (c,d). 
For the results of analogous analyses, see \cite{arn,cn,bot}.

 In addition we
show the constraint coming from $m_{H_2} > 113$ GeV for $A_0=0$ which restricts one to
relatively large neutralino masses. For the cases where $\mu > 0 $, there is a potential 
upper limit to the neutralino mass coming from the recent BNL E821 experiment
\cite{BNL}, which reports a new value for the anomalous magnetic moment
of the muon: $g_\mu - 2
\equiv 2 \times a_\mu$ that is apparently discrepant
with the Standard Model
prediction at the level of 2.6 $\sigma$.  This limit is also displayed on Figure 1
\cite{ENO,arn3}. For $\tan \beta = 10, \mu > 0$, the theory is quite predictive in both
the LSP mass and scattering cross section. No value of $\tan \beta$ is
compatible with the BNL E821 result if $\mu < 0$.

The scalar cross section is, in general, more
sensitive to the sign of $\mu$ than is the spin-dependent 
cross section.
Notice that, in Figure 1c for $\tan \beta = 10$ and $\mu < 0$, there is
a cancellation. Higgs exchange is dominant in
$\alpha_3$ and for $\mu < 0$, both $Z_{\chi 3}$ and
$Z_{\chi 4}$ are negative, as is
the Higgs mixing angle $\alpha$. 
Inserting the definitions of $\delta_{1i(2i)}$, we see that there
is a potential cancellation of the Higgs contribution to $\alpha_3$ for
both up-type and down-type quarks. Whilst there is such a cancellation
for the down-type terms, which change from positive to negative as
one increases
$m_\chi$, such a cancellation does not occur for
the up-type terms, which remain negative in the region of parameters we
consider. The cancellation
that is apparent in the figure is due to the cancellation in
$\alpha_3$ between the up-type contribution (which is negative) and the
down-type contribution, which is initially positive but
decreasing, eventually becoming negative as we increase $m_{\chi}$.

\begin{figure}
\begin{minipage}{8in}
\includegraphics[width=0.35\textwidth]{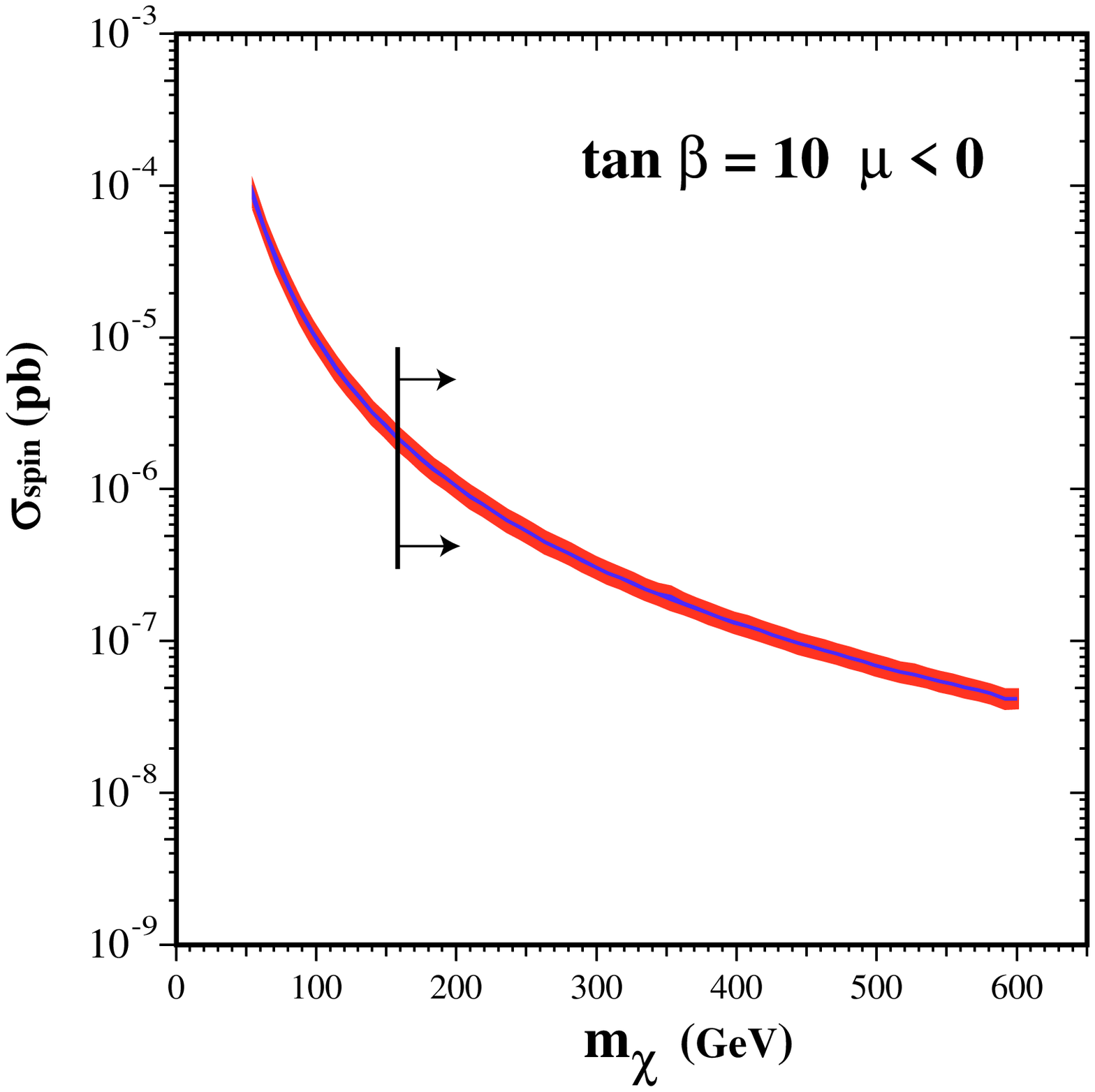}
\hspace*{-.3in}
\includegraphics[width=0.35\textwidth]{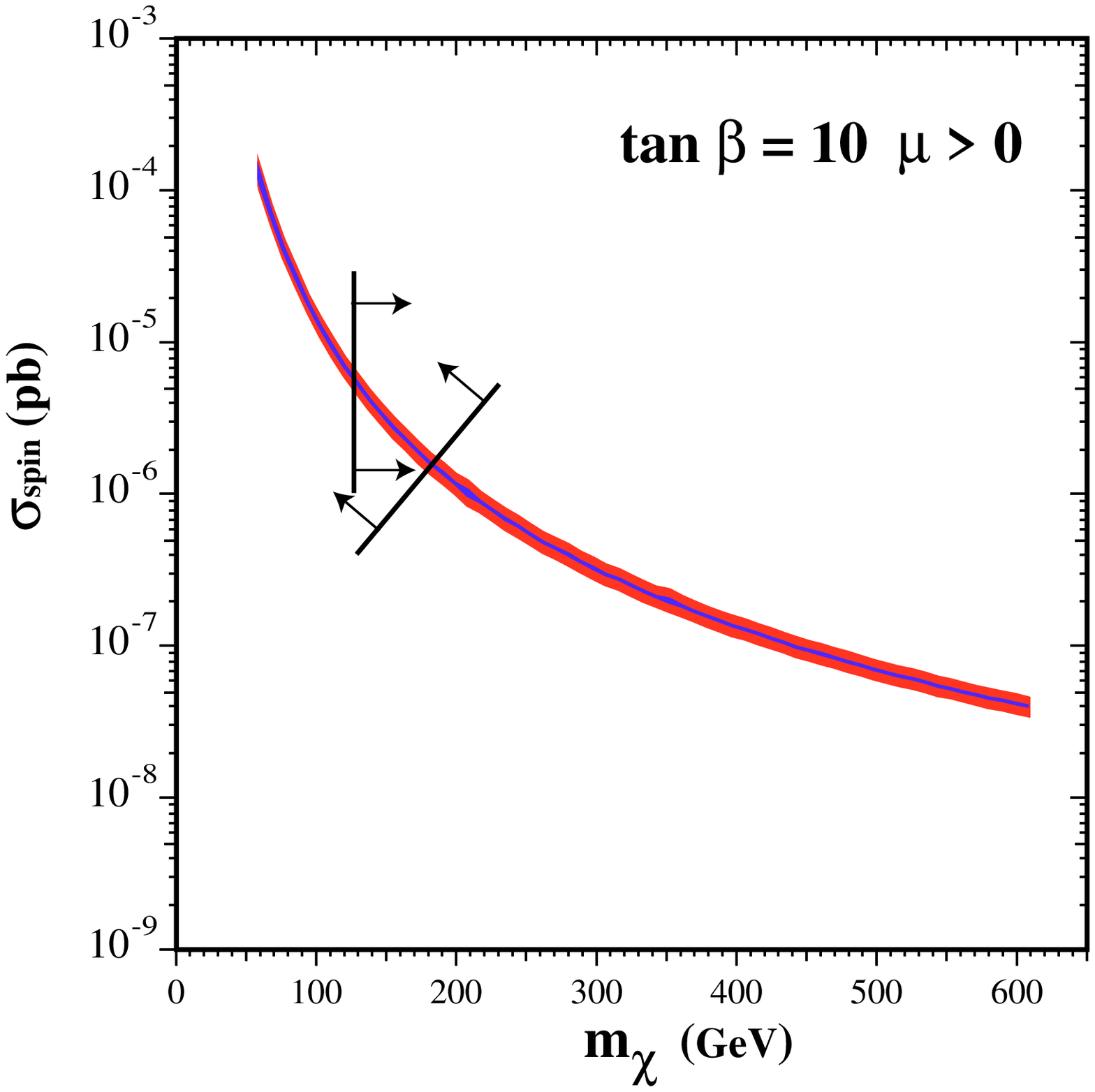}
\end{minipage}
\begin{minipage}{8in}
\vspace*{-1in}\includegraphics[width=0.35\textwidth]{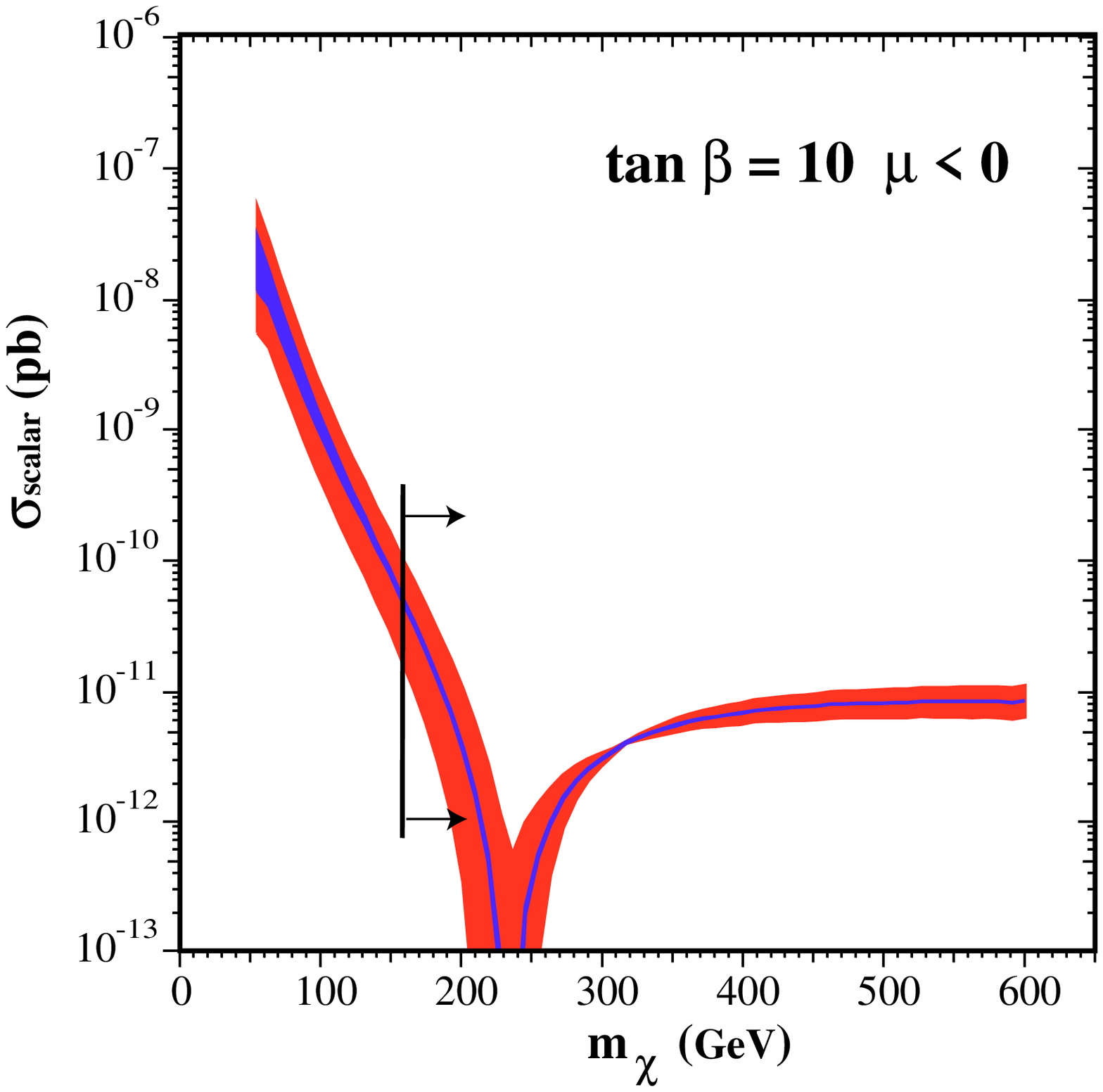}
\hspace*{-.3in}
\includegraphics[width=0.35\textwidth]{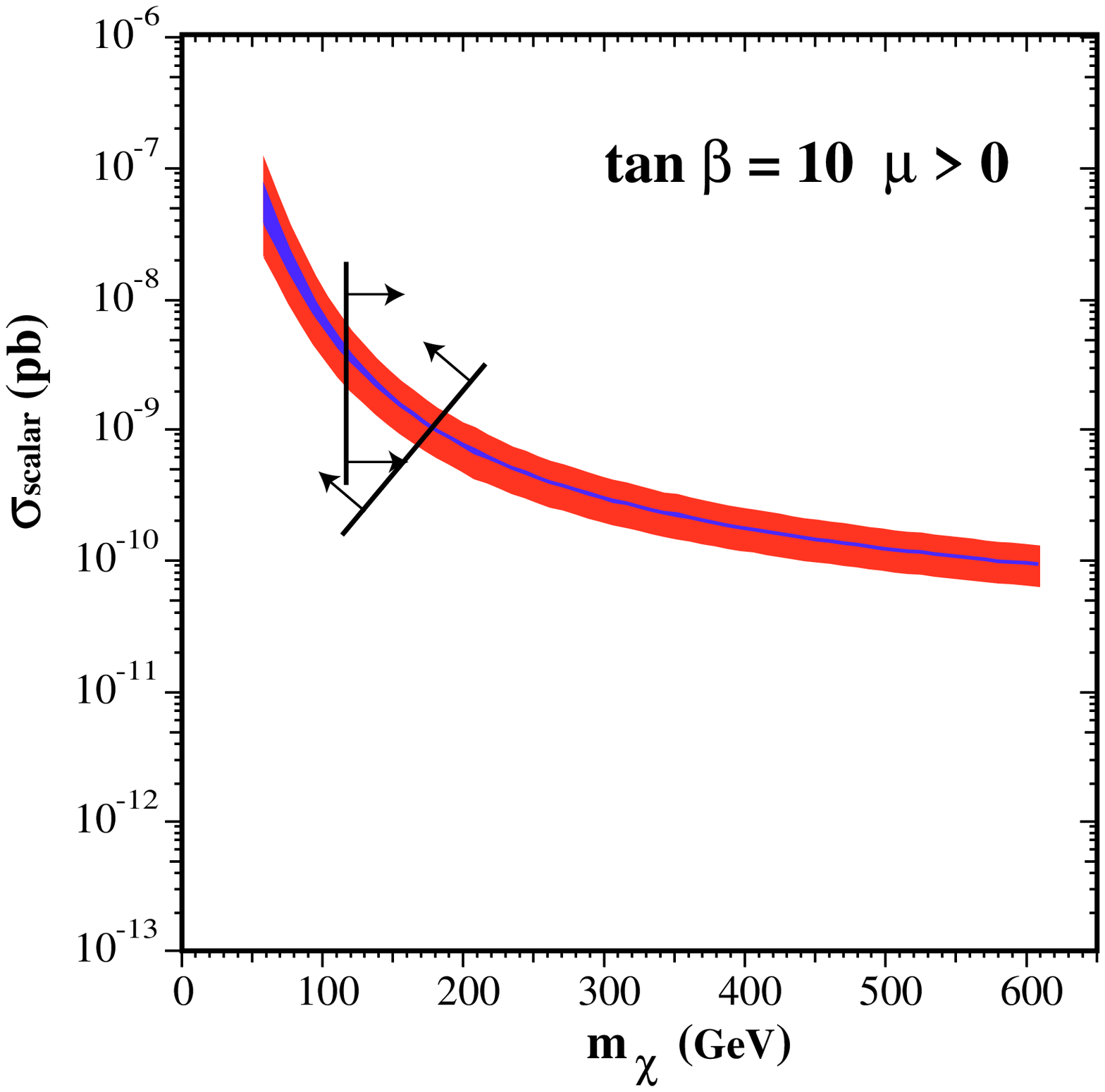}
\end{minipage}
\caption{\label{fig:spin}
{\it (a,b): The spin-dependent cross section for the elastic scattering of
neutralinos on protons as a function of the LSP mass for $\tan \beta = 10$.
The central curves are based on the inputs
(\protect\ref{spincontent}), and their thicknesses are related to
the spreads in the allowed values of $m_0$.  The shaded regions
correspond to the uncertainties in the hadronic inputs
(\protect\ref{spincontent}).
(c,d): The spin-independent scalar cross section for the elastic scattering
of neutralinos on protons as a function of the LSP mass for $\tan \beta = 10$. The
central curves are based on the inputs (\protect\ref{pinput}),
their thicknesses are again related to the spread in the
allowed values of $m_0$, and the shaded regions now correspond to the
uncertainties in the hadronic inputs (\protect\ref{pinput}). The
supplementary lower
limits imposed on $m_\chi$ in this and the next figure reflect
improvements in the LEP lower limit on $m_h$, and the upper
limits for $\mu > 0$ are due to $g_\mu - 2$, which is
incompatible with $\mu < 0$. }}
\end{figure}

At higher values of $\tan \beta$, one can in principle expect larger elastic cross
sections \cite{arn,lns}.  In Figure 2, we show the spin dependent
(a,b) and the scalar (c,d)  cross section for $\tan \beta = 35, \mu < 0$
(a,c) and for $\tan \beta = 50, \mu > 0$ (b,d).  In this figure, lower values of $m_\chi$
have been cut off (and are not shown) due to the constraint imposed by
measurements of $b \to s~\gamma$.

As was discussed in detail in \cite{EFGOSi}, a new feature in the $m_{1/2} - m_0$ plane
with acceptable relic density appears at large $\tan \beta$. At large $m_{1/2} \sim 1000$
GeV, it becomes possible for neutralinos to annihilate through s-channel $H_1$ or
pseudoscalar, $A$, exchange.  In fact there is a slice in the plane where
$2 m_\chi \approx m_{H_1,A}$ and the relic density becomes uninterestingly small.
At smaller and larger $m_{1/2}$ surrounding this pole region, there are regions
where the relic density falls in the desired range.  This leads to two separate regions
in Figure 2 at lower $m_\chi$.  The third region in Figure 2 at higher $m_\chi$
corresponds to the cosmological region allowed by coannihilation
\cite{EFOSi}. For more
further details on $H_1,A$-pole and coannihilation, see the contribution of John Ellis in
these proceedings. As in the case $\tan \beta = 10, \mu < 0$, the scalar cross
section at higher $\tan \beta$ also exhibits the cancellation feature discussed above.
However, because the cosmological regions are multivalued in $m_0$ as a
function of 
$m_{1/2}$, the cancellation occurs at a different value of $m_\chi$ for each of three
regions just discussed. This leads (unfortunately) to a broad region in the
$\sigma-m_\chi$ plane where the cross section is {\em very} small.

\begin{figure}
\begin{minipage}{8in}
\includegraphics[width=0.35\textwidth]{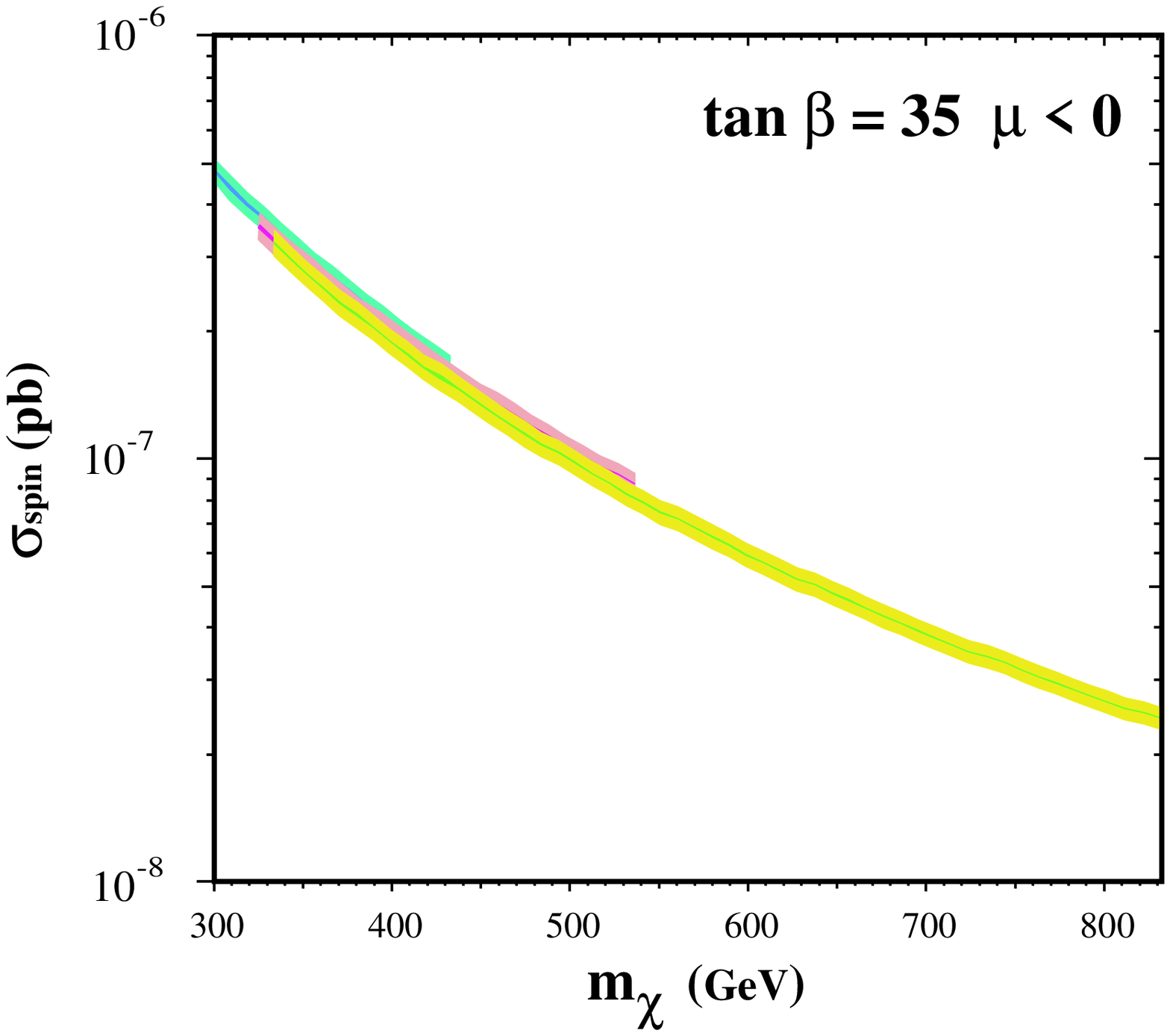}
\hspace*{-.3in}
\includegraphics[width=0.35\textwidth]{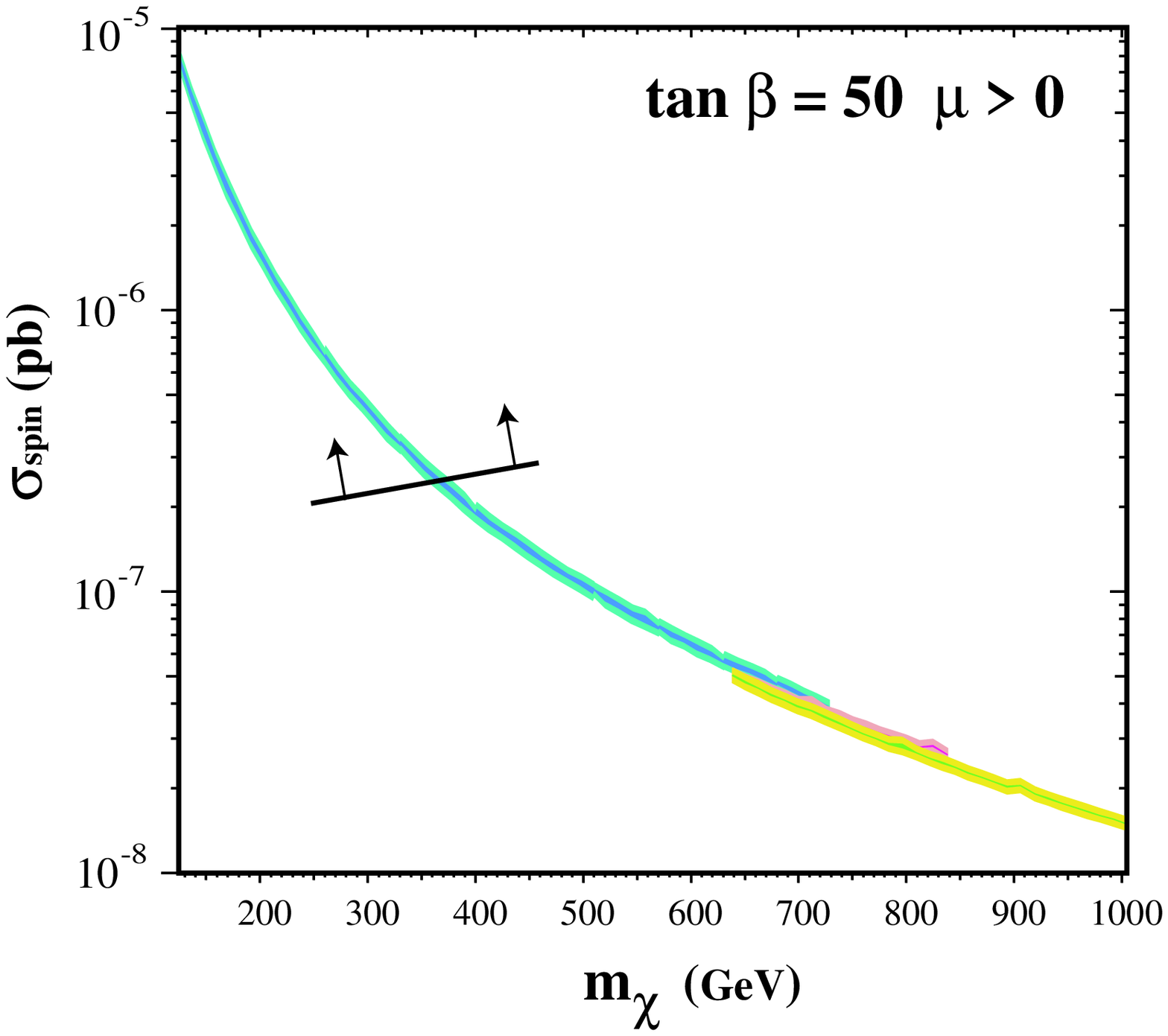}
\end{minipage}
\begin{minipage}{8in}
\vspace*{0in}\includegraphics[width=0.35\textwidth]{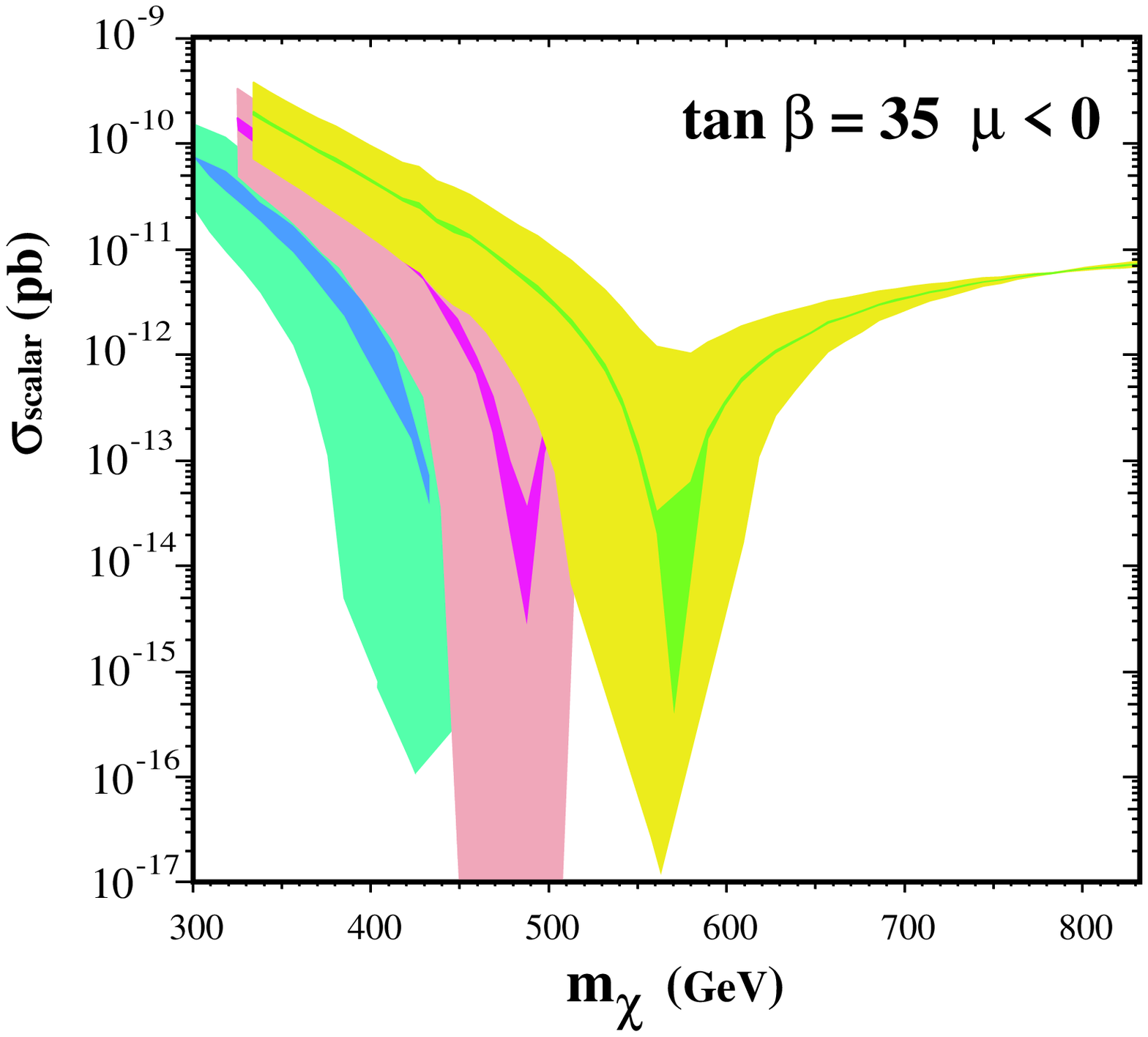}
\hspace*{-.3in}
\includegraphics[width=0.35\textwidth]{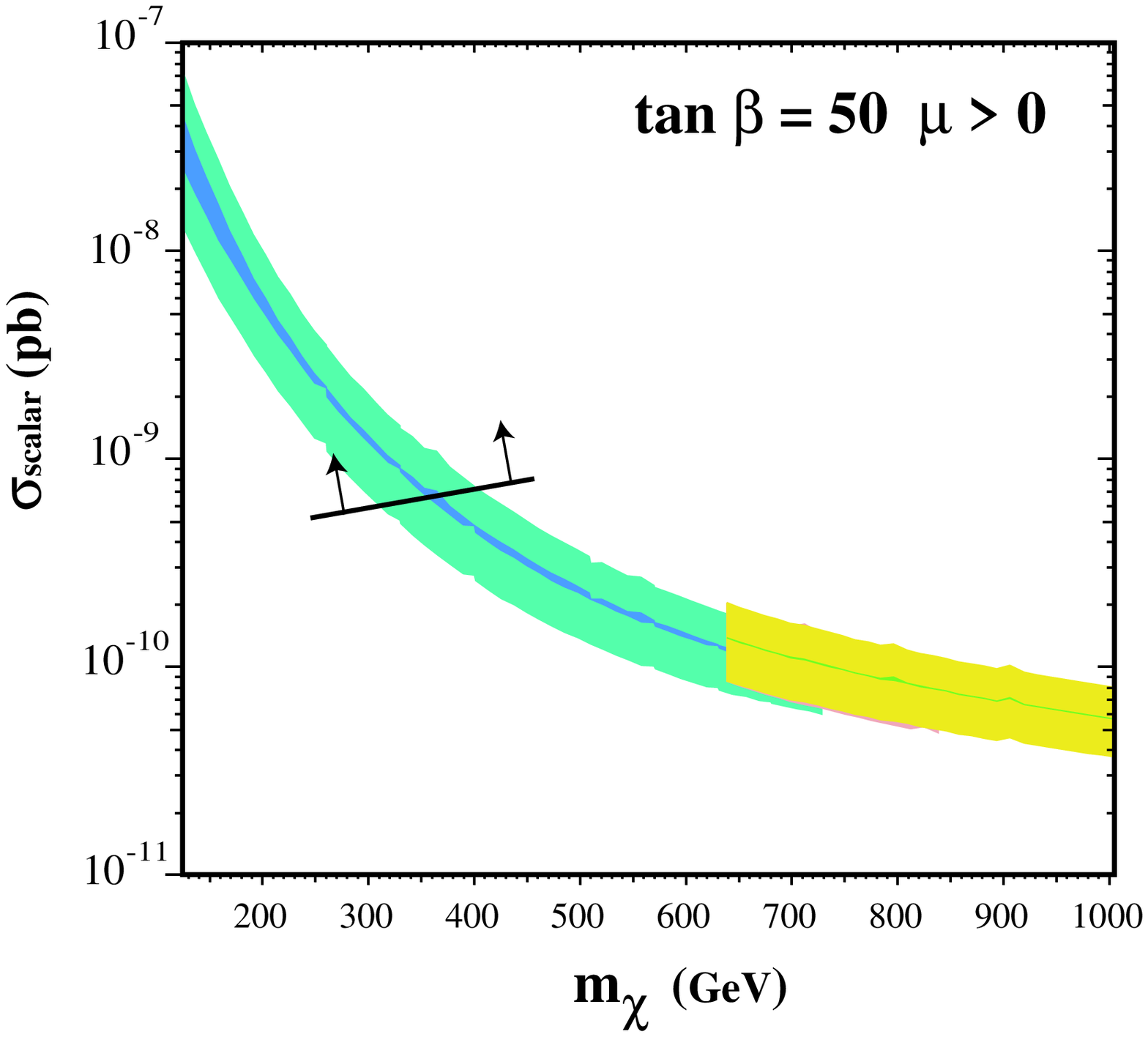}
\end{minipage}
\caption{\label{fig:spinh}
{\it As in Figure 1 for $\tan \beta = 35$.
Here the three distinct regions correspond to the two sides of the
$H_1,A$ annihilation poles, and the coannihilation region at higher
values of $m_\chi$. 
}}
\end{figure}

In the MSSM, in addition to scanning over the gaugino and sfermion masses at
fixed $\tan \beta$, one can treat $\mu$ and the pseudoscalar mass $m_A$
as free parameters as well.  In \cite{EFlO2}, we performed a scan over the following
parameter space:
$
0  < m_0 <  1000 ;
80  < |\mu| <  2000 ;
80  < M_2 <  1000 ;
0  < m_A <  1000 ;
-1000  < A <  1000.
$
Of the 90,000 (70,000) points scanned for $\tan \beta = 10$ and $\mu > 0$ ($\mu < 0$),
only 6208 (4772) survived all of the experimental and cosmological constraints.
In the CMSSM, the LSP is nearly always predicted to be Bino of very high purity.
However, in the MSSM, when $|\mu| \la M_2$, the LSP may have a dominant
Higgsino component. In these cases, coannihilation \cite{gs} greatly suppresses their
relic density and when combined with the experimental constraints on the parameter
space, Higgsino dark matter can be excluded as a viable option \cite{EFGOS,EFGO}.

The LEP chargino and Higgs cuts
remove many points with low $m_\chi$ and/or large elastic scattering
cross sections. The sfermion mass cut is less important.
The constraint that $\chi$ be the LSP removes quite a large number of
points, populated more or less evenly in the cross section plots.
There is a 
somewhat sparse set of points with very small cross sections which give
some measure of how low the cross section may fall in some special cases.
These reflect instances where particular cancellations take place,
as discussed above. The lower boundary of the densely occupied regions
 offers an answer to the question how low the elastic scattering cross sections may
reasonably fall, roughly
$\sigma \sim 10^{-9}$~pb for the spin-dependent cross section and
$\sim 10^{-10}$~pb for the spin-independent cross section.

Our resulting predictions for the spin-dependent elastic neutralino-proton
cross section for $\tan\beta = 10$ are shown in
Figure 3(a,b), where a comparison with the CMSSM is also
made.
The raggedness of the upper and lower boundaries of the dark (blue) shaded
allowed
region reflect the coarseness of our parameter scan, and the relatively
low density of parameter choices that yield cross sections close to
these boundaries. 
it should be noted that the low values of $m_\chi$ in these plots, that
yield relatively high
spin-dependent cross section, have now been excluded by
improvements in the Higgs mass limit.  As $m_\chi$ increases, the maximum
allowed
value of $\sigma_{spin}$ decreases, though not as rapidly as in the
previous CMSSM case~\cite{EFlO1}. The hadronic uncertainties are basically
negligible for this spin-dependent cross section, as seen from the
light (yellow) shading.

\begin{figure}
\begin{minipage}{8in}
\includegraphics[width=0.35\textwidth]{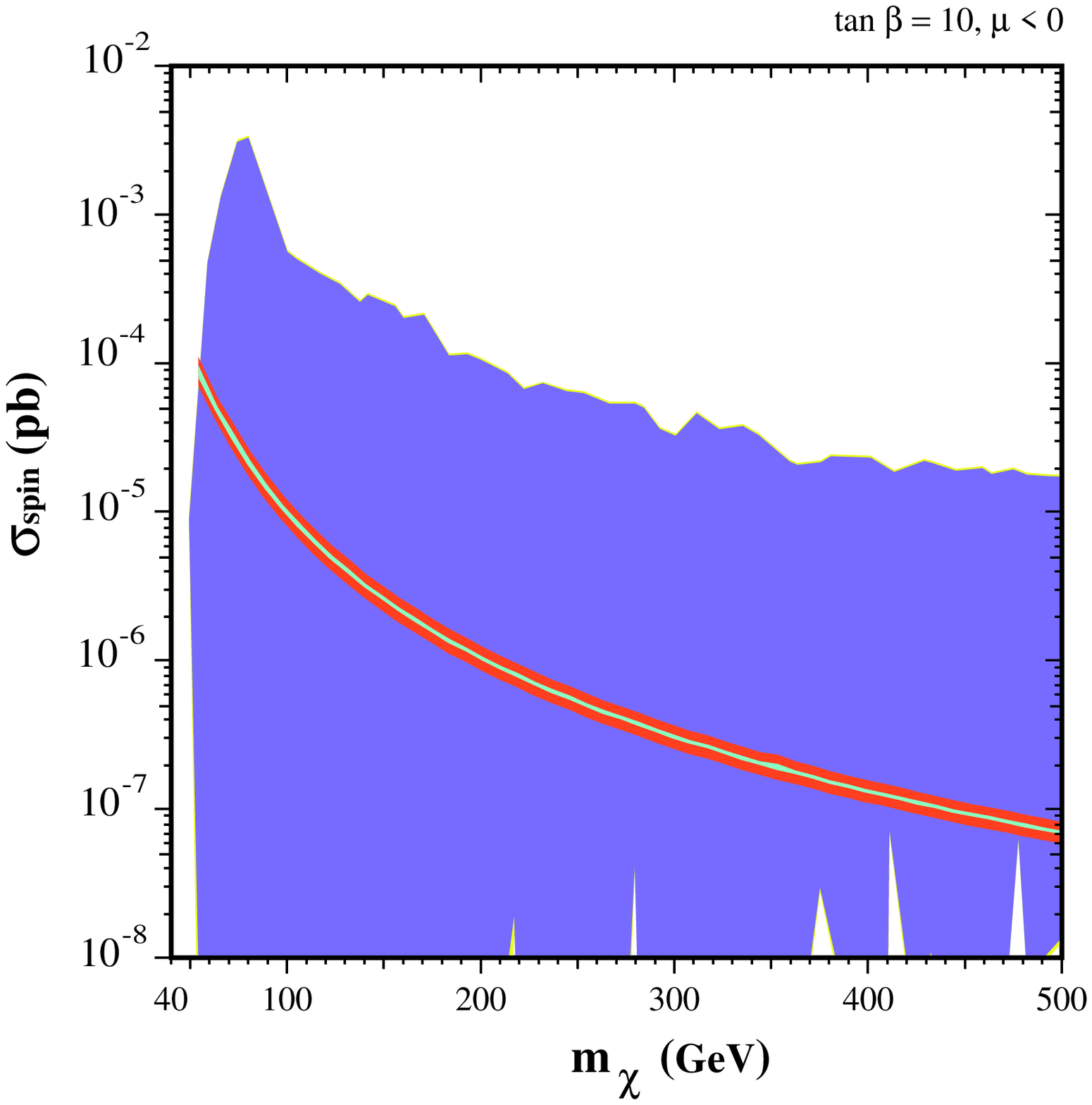}
\hspace*{-.3in}
\includegraphics[width=0.35\textwidth]{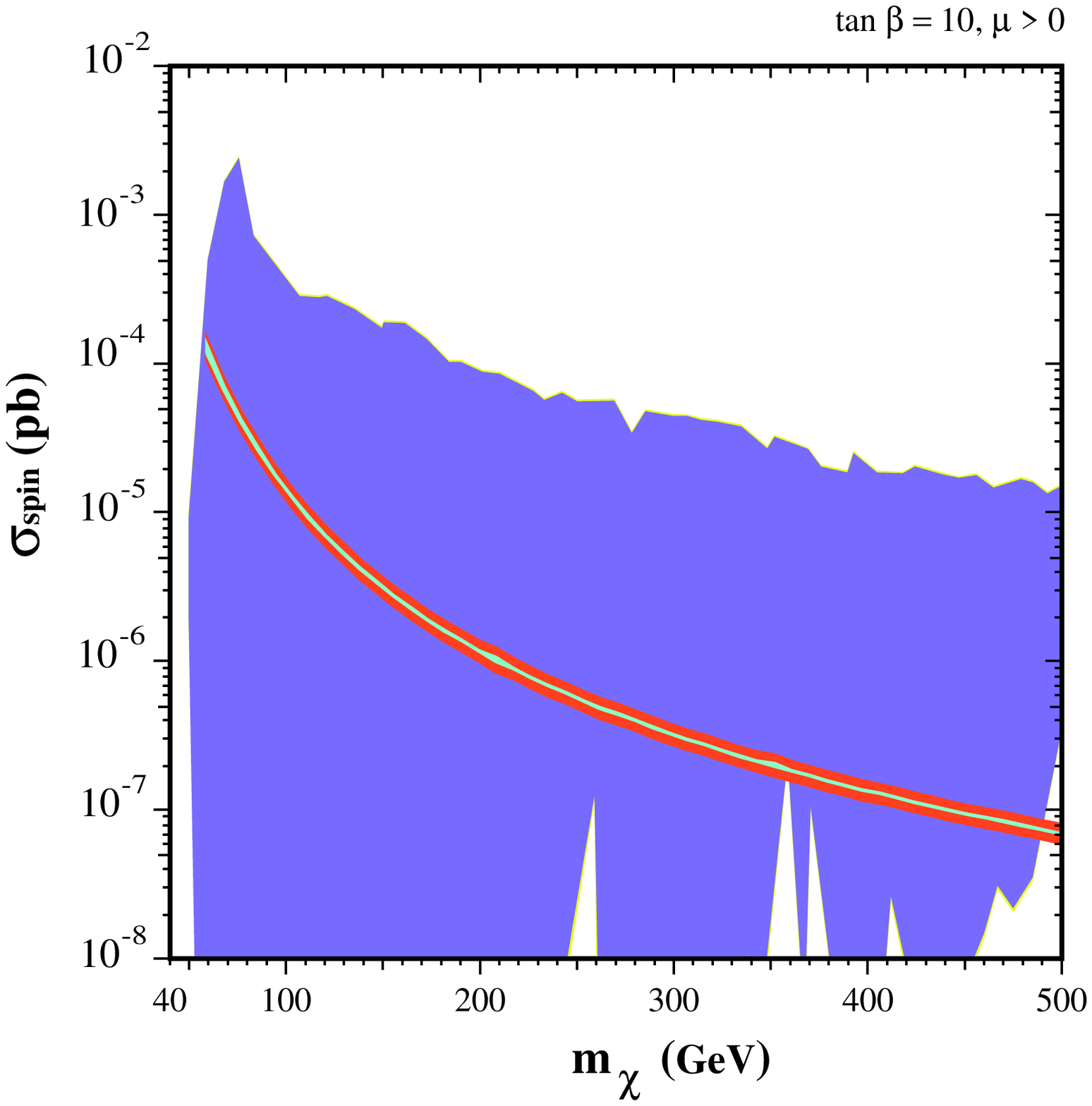}
\end{minipage}
\begin{minipage}{8in}
\vspace*{-1in}\includegraphics[width=0.35\textwidth]{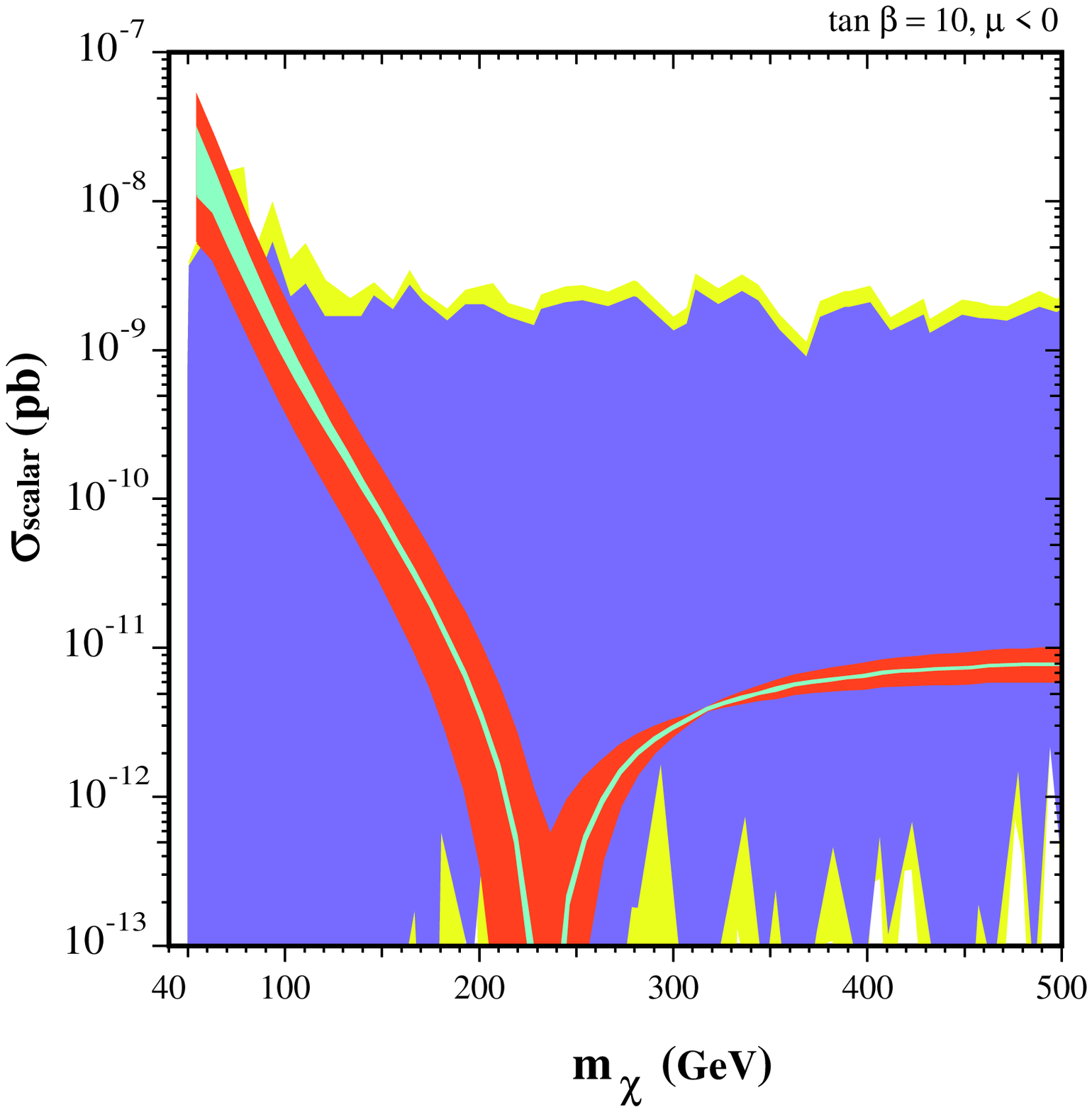}
\hspace*{-.3in}
\includegraphics[width=0.35\textwidth]{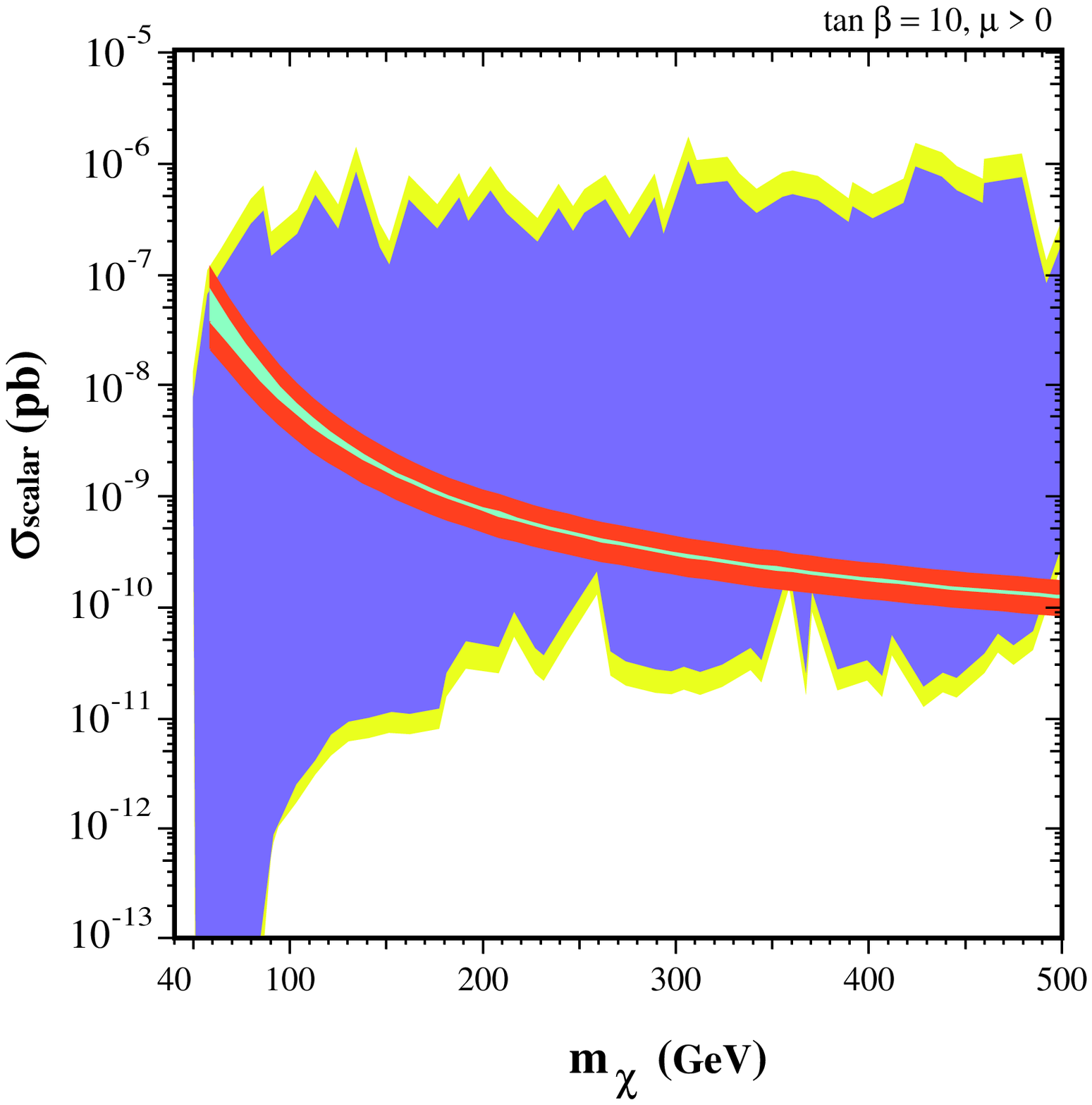}
\end{minipage}
\caption{\label{fig:spins}
{\it  As in Figure 1. The main (blue) shaded
regions summarize the envelopes of possible values found in our scan,
for points respecting the LEP constraints, discarding points with
$\Omega_\chi h^2 > 0.3$, and rescaling points with $\Omega_\chi h^2 <
0.1$. 
The small light (yellow) shaded
extensions of this region reflect the hadronic matrix element
uncertainties. The concave
(red and turquoise) strips are those
found previously assuming universal Higgs scalar masses \protect\cite{EFlO1}.
}}
\end{figure}

The analogous results for the spin-independent elastic
neutralino-proton cross section are shown in  
Figure 3(c,d), where comparisons with the CMSSM case are also
made.
We see a pattern
that is similar to the spin-dependent case.
For small $m_\chi$, the spin-independent
scalar cross section, shown by the dark (blue) shaded region, may be
somewhat higher than in the CMSSM case, shown by the (red and turquoise)
diagonal strip, whilst it could be much smaller. For large $m_\chi$,
the cross section may be rather larger than in the CMSSM case, but it is
always far below the present experimental sensitivity.  
Overall, we note that the hadronic uncertainties, denoted by the light
(yellow) bands, are somewhat larger in the spin-independent case
than in the spin-dependent case.

\section{Conclusions}
\label{sec:conclusions}

As one can see from scanning the figures, the predicted elastic scattering 
cross section in the CMSSM and in the more general MSSM, are relatively small.
For the spin-dependent processes, the cross sections fall in the range
$\sigma \sim 10^{-4} - 10^{-8}$ pb, whereas for the scalar cross sections,
we find $\sigma < 10^{-6}$  pb with an uncertain lower limit
due to possible cancellations. These should be compared with
current sensitivities of existing and future experiments
\cite{gm}. The UKDMC detector is sensitive to $\sigma \ga 0.5$ pb for the spin-dependent
cross section. DAMA and CDMS are sensitive to $\sigma \ga 2 \times
10^{-6}$ pb for the
scalar cross section.  This is close to the upper limits we find for reasonable
supersymmetric models.  The future looks significantly brighter.
When CDMS is moved to the Soudan mine, its sensitivity will drop to
between
$10^{-8}$ and $10^{-7}$ pb and GENIUS claims to be able to reach $10^{-9}$pb.
At those levels, direct detection experiments will 
either discover supersymmetric dark matter or impose serious 
constraints on supersymmetric models.

\acknowledgements
The work of K.A.O. was supported in part by DOE grant
DE--FG02--94ER--40823.

\bibliographystyle{klunamed}
{}
\end{article}

\end{document}